\documentclass[twocolumn,showkeys,preprintnumbers,amsmath,amssymb,aps]{revtex4}

\usepackage{graphicx}
\usepackage{dcolumn}
\usepackage{bm}
\usepackage{epsfig}

\begin{document}


\title{Terminal stage of highly viscous flow}

\author{U. Buchenau}
 \email{buchenau-juelich@t-online.de}
\affiliation{%
Forschungszentrum J\"ulich GmbH, J\"ulich Centre for Neutron Science (JCNS-1) and Institute for Complex Systems (ICS-1),  52425 J\"ulich, GERMANY}%

\date{May 8, 2022}

\begin{abstract}
The shear misfit model for the highly viscous flow is based upon a theoretical prediction for its terminal stage in terms of irreversible Eshelby relaxations in the five-dimensional shear space. The model is shown to predict a small $\delta$-function (Debye peak) in the dielectric spectrum, in agreement with experimental evidence. It is extended from shear fluctuations to density fluctuations, and new relations between adiabatic and isothermal compressibility jumps at the glass transition are derived. The model is applied to high precision measurements of the shear, dielectric and bulk relaxation data in two vacuum pump oils and in squalane, a short chain polymer with a strong secondary relaxation peak. In all three substances, the adiabatic compressibility equilibrates much earlier than the isothermal one, due to the faster equilibration of the vibrational entropy. The terminal stage of aging data in squalane demonstrate that one sees also the adiabatic density fluctuations in the thermal expansion, explaining why it seems to equilibrate a bit faster than the dynamic heat capacity. The Prigogine-Defay ratio at the glass transition finds a new explanation in terms of different Gr\"uneisen parameters of the boson peak modes and higher frequency vibrations. 
\end{abstract}

\maketitle 

\section{Introduction}

The highly viscous flow of glass formers close to the glass transition \cite{jackle,cavagna,bb,royall,berthier} is still a fighting ground for different theoretical approaches, some of them derived from liquid theory, but also an increasing number of concepts \cite{shov,falk,johnson,ngai,dyre,random,bu2018a,bu2018b} which attribute the highly viscous flow to thermally activated structural relaxations in the glass. So far, none of these explanations links the glass transition properties to the low temperature glass anomalies \cite{ramos,schober}, the tunneling states and the soft vibrational modes at the boson peak, which are at present the subject of intense numerical studies \cite{bouch}. 

Among the glass relaxation concepts \cite{shov,falk,johnson,ngai,dyre,random,bu2018a,bu2018b}, there is a theoretical analysis \cite{bu2018a} of the terminal stage of the highly viscous flow. According to the analysis, the terminal stage is due to irreversible shear transformation processes in the five-dimensional shear space in asymmetric double-well potentials, with the asymmetry determined by the different shear misfits of the inner Eshelby domain \cite{eshelby} or shear transformation zone \cite{falk,johnson}, with respect to the surrounding viscoelastic matrix, in its two structural alternatives. One finds an Eshelby region lifetime $\tau_c=8\eta/G$ ($\eta$ viscosity, $G$ short time shear modulus), eight times longer than the Maxwell time, derived from the Einstein relation between the irreversible shear fluctuations and the viscosity.

There is a pragmatical extension of the theoretical analysis of irreversible Eshelby relaxations to the reversible Eshelby relaxations at shorter relaxation times \cite{bu2018b}, taking both reversible and irreversible processes to belong to the same Kohlrausch distribution, the irreversible ones for relaxation times longer than $\tau_c$, and the reversible ones for relaxation times shorter than $\tau_c$. The extension, here denoted as shear misfit model, is able to describe the shear relaxation of simple glass formers without strong hydrogen bonds and without secondary relaxation in terms of only three parameters, namely $G$, $\tau_c$, and the Kohlrausch exponent $\beta$ close to 1/2. It holds the promise to be developed into a valid theory of the highly viscous flow, hopefully including an understanding of the soft vibrational modes \cite{ramos,schober,bouch}.

The present paper begins with the further development of the shear misfit model, showing that it predicts a small $\delta$-function in dielectric and other spectra. The second aim is the quantitative treatment of density fluctuations, deriving their relation to shear and energy fluctuations. Thirdly, for glass formers with secondary relaxations, one can reduce its number of temperature-dependent parameters to three, namely the short time shear modulus, the terminal relaxation time $\tau_c$, and the amplitude of the secondary relaxation peak. The model is applied to measurements in two vacuum pump oils \cite{tina} and in squalane, a short chain polymer with a strong secondary relaxation peak, where there is a happy combination of high precision shear relaxation measurements \cite{nibo,squa2017} with a high precision terminal aging measurement \cite{ag2020}.

After this introduction, the paper proceeds with the discussion of the previous theoretical shear misfit work \cite{bu2018a,bu2018b} in Section II, correcting three smaller errors in the former papers, and deriving equations for adiabatic and isothermal density fluctuations. Section III describes the fit of shear \cite{tina,nibo,squa2017}, compressibility \cite{tina,hex}, dielectric \cite{tina,nibo}, and aging data \cite{ag2020} in two vacuum pump oils (DC704 and PPE) and squalane in terms of the shear misfit model. Section IV discusses and summarizes the results.

\section{Theoretical basis}

\subsection{Irreversible Eshelby relaxations}  

Consider a more or less spherical region of ten to hundred atoms or molecules in the undercooled liquid. It is reasonable to define a dimensionless shear misfit $e$ of the region in such a way that $e^2$ is the shear misfit energy in units of $k_BT$. According to Eshelby's theoretical treatment of this situation \cite{eshelby}, about half of the elastic distortion energy is a shear distortion of the region itself; the other half is a more complicated elastic distortion of the outside, with a large shear component and a small bulk compression one.

In thermal equilibrium, the states $e$ in the five-dimensional shear misfit space have an average energy of 5/2 $k_BT$ in the normalized distribution
\begin{equation}\label{pe}
	p(e)=\frac{1}{\pi^{5/2}}e^4\exp(-e^2).
\end{equation}
The prefactor corrects the one of eq. (3) in the previous paper \cite{bu2018a}, a mistake which does not invalidate the results.

The cornerstone of the shear misfit model is the analysis \cite{bu2018a} of the terminal part of the flow process in terms of irreversible Eshelby region transformations (shear transformation zones) \cite{eshelby,falk,johnson}, which change the elastic shear misfit of the Eshelby region. Assuming a constant density of stable structural solutions in the five-dimensional shear space, and an Einstein relation between the irreversible shear fluctuations and the viscosity $\eta$, one finds the terminal relaxation time 
\begin{equation}\label{tauc}
	\tau_c=8\tau_M=\frac{8\eta}{G},
\end{equation}
where $G$ is the short time shear modulus, and $\tau_M$ is the Maxwell time.

The lifetime $\tau$ of a given Eshelby region depends strongly on its elastic shear misfit energy $e^2$. If this is high, there are many possibilities for an irreversible decay into a structure with a smaller elastic shear misfit energy. Integrating over the states in the distribution of eq. (\ref{pe}) in terms of the barrier variable $v=\ln(\tau/\tau_c)$, one finds \cite{bu2018a} the normalized irreversible spectrum
\begin{equation}\label{pt}
	l_{irrev}(v)=\frac{1}{3\sqrt{2\pi}}\exp(2v)\left(\ln(4\sqrt{2})-v\right)^{3/2},
\end{equation}
a slightly broadened Debye process around the relaxation time 1.75 $\tau_c$, a factor of fourteen longer than the Maxwell time. In fact, the corresponding decay function is well fitted by the Kohlrausch function $\exp(-(t/1.8\tau_c)^{0.8}$. Eq. (\ref{pt}) is consistent with eq. (16) of ref. \cite{bu2018a}, and corrects eq. (10) of ref. \cite{bu2018b}, again a mistake which does not invalidate the results of this second paper, because the correct equation was used in the data evaluation.

In the derivation of eq. (\ref{pt}), three effects are neglected. The first is the time dependence of the energy $e^2$ arising from the other Eshelby processes in the neighborhood, which leads to a diffusive motion of $e^2$. This effect causes a narrowing of the spectrum, because a slowly decaying low shear energy Eshelby region is able to drift toward faster decay times, and the opposite happens on the high energy side.

The second neglected effect is the barrier distribution of the irreversible processes, which was assumed to be a $\delta$-function in the derivation of eq. (\ref{pt}). In the treatment of the reversible Eshelby relaxations in the next subsection II. B, it will be seen that there is good reason to assume a barrier distribution with a rather large finite width, which ought to lead to a broadening of the spectrum.

The third neglected effect is the volume change of the Eshelby region in its irreversible transitions (responsible for the terminal aging data of the density in squalane \cite{ag2020} used in Section III. B to corroborate the shear misfit model).

But these three effects do not affect the predicted position. Since the irreversible Eshelby decay spectrum of eq. (\ref{pt}) is able to describe not only the position, but also the width of measured dynamic heat capacity spectra in four glass formers with Maxwell times determined from shear data at the same temperature \cite{bu2018a,bu2018b}, it seems probable that the three neglected effects cancel.  

The average decay rate $\overline{r}=\overline{1/\tau}$ of the distribution of eq. (\ref{pt}) is $1/\tau_c$. In this sense, $\tau_c$ can be considered as the average lifetime of both irreversible and reversible Eshelby relaxations, and the appropriate cutoff function for the reversible Eshelby relaxations is the Kohlrausch function $\exp(-(\tau/1.8\tau_c)^{0.8})$, a result which is central for the description of the reversible relaxations.

\subsection{Reversible relaxations: Kohlrausch tail and secondary relaxations}

One needs an additional postulate \cite{bu2018b} to include the reversible Eshelby transformations, in the simplest case a Kohlrausch barrier density proportional to $\exp(\beta v)$ in terms of the barrier variable $v=\ln(\tau/\tau_c)$, with a Kohlrausch exponent $\beta$ close to 1/2 \cite{bnap,albena}. The postulate is that the Kohlrausch barrier density of the reversible relaxations extends without discontinuity or change of slope to barriers with $v>0$, which are irreversible transitions and are responsible for the viscous flow. Their flow contribution does not diverge, because the increase with $\exp(\beta v)$ is overcompensated by the rate factor $1/\tau=1/\exp(v)\tau_c$.

As already mentioned in II. A, this implies a barrier distribution proportional to $\exp(-v/2)(1-\exp(-(\exp(v-\ln{1.8}))^{0.8})$ for the irreversible processes, which is not a $\delta$-function, but rather one with a full width at half maximum of about four in $v$, corresponding to a width between one and two decades in relaxation times. But the resulting broadening of the irreversible spectrum of eq. (\ref{pt}) might still be small, because in the derivation of the equation the lifetime of a single Eshelby region is an integral over all possible Eshelby transitions.

A reversible relaxation has a factor 0.4409/2=0.22045 weaker contribution to the shear compliance than an irreversible one \cite{bu2018b}, due to two effects. The factor of 2 reflects the fact that the average shear and compression stress energy of the surroundings does not disappear in a reversible relaxation. The factor 0.4409 stems from the different effects of the energy asymmetry $\Delta=e_0^2-e^2$ between the initial and the final state of a structural relaxation for reversible and irreversible relaxations, integrated over all possible combinations \cite{bu2018b}.

One can describe the reversible shear relaxation processes of simple glass formers without secondary relaxation peaks in terms of the Kohlrausch barrier density
\begin{equation} \label{lv}
	l_{rev}(v)=f_0\exp(\beta v)F_c(v),
\end{equation}
with the Kohlrausch cutoff function
\begin{equation}
	F_c(v)=\exp(-(\exp(v-\ln{1.8})^{0.8})
\end{equation}
and $f_0$ given by
\begin{equation}\label{f0}
	f_0=0.4409\frac{8-4\beta}{3}.
\end{equation}

This prefactor $f_0$ ensures that the irreversible Eshelby transitions under an external shear stress $\sigma$ lead to a viscous shear strain displacement of $8\sigma/G$ after the time $\tau_c$, taking their enhancement factor 2/0.4409 with respect to the reversible transitions into account. In simple cases, glass formers without strong hydrogen bonds (the hydrogen bonded glass formers are treated in a separate paper \cite{hb}) and with no secondary relaxation peak \cite{bu2018b}, eq. (\ref{f0}) works, and the whole shear relaxation is described by the three parameters $G$, $\tau_c$ and $\beta$, with $\beta$ close to 1/2 in all investigated cases. No other shear relaxation model is able to do that.

If one has a secondary shear relaxation peak from changes of the shape or the orientation of the molecule, these shape or orientation changes do not contribute to the viscosity. The viscous flow requires an irreversible change of the molecular packing, while changes of the molecular shape or orientation are always reversible in the long run. The gain in shear compliance by a change of the shape or the orientation of the molecule in an irreversible Eshelby relaxation is lost, when the molecule returns in one of the subsequent irreversible relaxations to its former shape or orientation.

In these cases, one has to add an appropriate gaussian distribution $l_G(v)$ to the Kohlrausch barrier density to describe the secondary relaxation peak \cite{gainaru}
\begin{equation} \label{lvs}
	l_{rev}(v)=(f_K\exp(\beta v)+l_G(v))F_c(v),
\end{equation}
introducing four more fit parameters, the enhancement factor $f_K$ for the Kohlrausch barrier density, the amplitude $a_G$, position $v_G$ and width $\sigma_G$ of the gaussian
\begin{equation}
	l_G(v)=a_G\exp((v-v_G)^2/2\sigma_G^2).
\end{equation}
The position $v_G$ is related to the central barrier $V_G$ of the gaussian by
\begin{equation}
	v_G=\frac{V_G}{k_BT}-\ln{\tau_c}+\ln{\tau_0},
\end{equation}
where $\tau_0$ is the microscopic relaxation time of a thermally activated process between 10$^{-13}$ and 10$^{-14}$ s.

The enhancement factor $f_K$ is necessary, because one expects an admixture of secondary relaxation processes to the irreversible Eshelby relaxations proportional to their integral $I=\int l_G(v)dv$, an admixture which does not contribute to the viscous flow, and thus requires a higher $f_K$. The integral over the reversible Kohlrausch tail needed for the flow is 2.1 for $\beta=1/2$, so from the ratio of the two integrals over the reversible processes, one expects the enhancement factor $1+I/2.1$.

In the two examples with secondary relaxation peaks evaluated in terms of the shear misfit model \cite{bu2018b} one finds indeed a temperature-independent enhancement factor $f_s$
\begin{equation}\label{fks}
	f_K=1+f_sI,
\end{equation}
but $f_s$ is larger than 1/2.1. In squalane, $f_s=0.9$, as demonstrated again in the new evaluation of the squalane shear relaxation data in Section III. A, and in dibutyl phtalate $f_s=4$ (the factors are not equal for the two substances, as stated erroneously in eq. (17) of reference \cite{bu2018b}).

With eq. (\ref{fks}), a glass former with a secondary relaxation peak has the seven free parameters $G$, $\tau_c$, $\beta$, $f_s$ and the three parameters of the gaussian. Of these seven parameters, only $G$, $\tau_c$, and the secondary peak amplitude $a_G$ are temperature-dependent. The position and width of the secondary relaxation peak can be determined from a measurement in the glass phase.

Having defined $l_{rev}(v)$, one can calculate the complex shear compliance $J(\omega)$ from
\begin{equation}\label{jom}
	GJ(\omega)=1+\int_{-\infty}^\infty \frac{l_{rev}(v)dv}{1+i\omega\tau_c\exp(v)}-\frac{i}{\omega\tau_M},
\end{equation}
and invert it to get $G(\omega)$. At $\omega=0$, the integral in eq. (\ref{jom}) is equal to $GJ_0-1$, where $J_0$ is the total recoverable compliance.

\subsection{Full relaxation spectrum, energy and density fluctuations}

The normalized full spectrum of all Eshelby relaxations, reversible and irreversible, is given by
\begin{equation}\label{tot}
	l_{tot}(v)=\frac{1}{8+GJ_0-1}(8l_{irrev}(v)+l_{rev}(v)/f_K).
\end{equation}

As pointed out in the second theoretical paper \cite{bu2018b}, the simultaneous knowledge of irreversible and reversible relaxation processes from the shear data implies the knowledge of all Eshelby shear relaxation processes of the substance, and enables one to compare with whatever one sees in other relaxation techniques. But to do that properly, one needs as much additional knowledge as possible.

Dynamic heat capacity measurements seem to see mainly the irreversible processes. The argument is supported by measurements of the dynamic heat capacity in four glass formers which are well described by eq. (\ref{pt}) alone, with $\tau_c$-values determined from shear relaxation data in the same substances \cite{bu2018a,bu2018b}. The additional knowledge to be taken into account here is that the excess heat capacity which equilibrates consists of a larger structural part and a smaller vibrational one, due to the soft vibrational modes \cite{ramos,schober,bouch}.

One important piece of information from the numerical work on the soft modes \cite{bouch} is their unstable core \cite{corei,corein}, stabilized on a saddle point in energy by the surrounding elastic medium. The inner region of a reversible Eshelby relaxation, with its large energy barrier of about 30 $k_BT_g$, will contain many such unstable cores. These soft modes have a strong coupling to an external shear \cite{ramos,schober,bouch}, and must therefore be strongly affected even by a reversible Eshelby transition. As a consequence, a reversible Eshelby transition is able to adapt its vibrational entropy to an external temperature change.  

The comparison of different relaxation times in the two vacuum pump oils DC704 and PPE \cite{bo} shows that the dielectric polarization and the adiabatic compressibility equilibrate both half a decade earlier than the dynamic heat capacity and the dynamic thermal expansion. In terms of the shear misfit model, this implies that the reversible Eshelby relaxations, which carry only the reversible quarter of the total shear relaxation, are responsible for most of the dielectric and the adiabatic compressibility equilibration. The appropriate description turns out to be an exponential decay of the dielectric polarization or the density deviation, with the time constant $\tau_d$. This exponential decay implies an exponential cutoff $F_d(v)$ of the dielectric or adiabatic compressibility spectrum
\begin{equation}
	F_d(v)=\exp(-\tau/\tau_d)=\exp(-\exp(v-v_d)),
\end{equation}
with $v_d=\ln(\tau_d/\tau_c)$. This cutoff function has been already considered in the earlier work \cite{bu2018b}.

But what was neglected in the earlier work, was the effect of the second irreversible decay of the Eshelby regions, which occurs after the time 2$\tau_c$. The quantity which decays with the time constant $\tau_d$, at this time still has a fraction $\exp(-2\tau_c/\tau_d)$ of its initial value, but then continues decaying with $\tau_d$ by the collective effect of all subsequent irreversible Eshelby relaxations. This "viscous" fraction must appear in the spectrum as a $\delta$-function at the relaxation time $\tau_d$. Therefore the normalized spectrum is
\begin{equation} \label{ld}
	l_{d}(v)=l_0l_{tot}(v)F_d(v)+a\delta(v-v_d)
\end{equation}
with
\begin{equation}\label{a}
	a=\exp(-2\tau_c/\tau_d)
\end{equation}
(in the adiabatic compressibility case possibly $a=0$, as argued below)
and the normalization condition
\begin{equation}
	\int_{-\infty}^\infty l_0l_{tot}(v)F_d(v)dv=1-a.
\end{equation}

The adiabatic compressibility spectrum is due to Eshelby transitions which change the volume $V$ of the Eshelby region. Such a volume change $\Delta V/V$ is opposed by the shear resistance of the surroundings (the situation considered in the shoving model \cite{shov}), leading to a diminution $c=\delta V/V$ of the volume change. The resulting distortion energy is $2GVc^2/3$ outside, and $BVc^2/2$ inside, where $B$ is the short time bulk modulus. The force balance leads to the diminution factor $3B/(3B+4G)$ for $\Delta V/V$. This reduction factor is exactly the one between adiabatic and isothermal compressibility contributions $\Delta\kappa_{ad}$ and $\Delta\kappa_{iso}$ from the structural relaxation, because the outer shear force disappears in the terminal stage of the flow process
\begin{equation}\label{dkap}
	\frac{\Delta\kappa_{ad}}{\Delta\kappa_{iso}}=\frac{3B}{3B+4G},
\end{equation}
where $B$ and $G$ are the glass moduli.

The adiabatic compressibility case differs from the dielectric or the dynamic depolarized scattering one, because the existence of the additional term in the isothermal compressibility implies an oscillating temperature term at low frequency, together with the oscillating pressure. At very low frequency, the oscillating temperature term compensates the pressure effect. Therefore in this case $a=0$ in eq. (\ref{ld}) cannot be excluded.

\section{Comparison to experiment}

\subsection{Shear, bulk and dielectric relaxation}

\begin{figure}[b]
\hspace{-0cm} \vspace{0cm} \epsfig{file=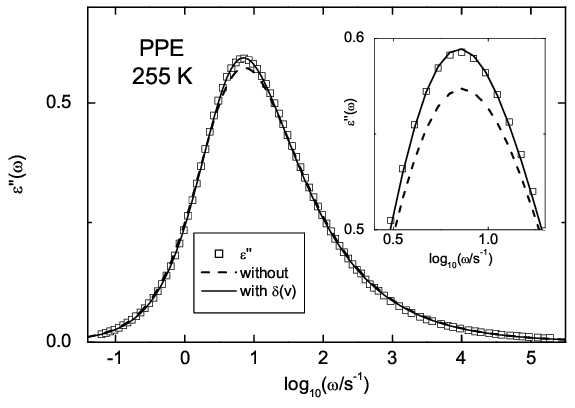,width=6 cm,angle=0} \vspace{0cm} \caption{Fit of dielectric data at 255 K in PPE \cite{tina} with (continuous line) and without (dashed line) the Debye peak from the irreversible processes. The insert shows the dashed line misfit at the peak.}
\end{figure}

The existence of a small $\delta$-peak in every dielectric glass former spectrum, a true Debye peak in terms of Debye's original meaning, has never been realized in the community, though it is generally acknowledged for the special case of monoalcohols \cite{mono}. However, with equs. (\ref{jom}) and (\ref{ld}), it is possible to search for it in experiment also in other glass formers.

In order to do this, one needs measurements of the shear relaxation and the dielectric relaxation of the same sample at the same temperature. Fitting the shear relaxation in terms of eq. (\ref{jom}), one gets the Eshelby region lifetime $\tau_c$ and the Kohlrausch $\beta$. According to eq. (\ref{ld}), that leaves only $\Delta\epsilon$ (the difference between low and high frequency dielectric constant, which can be also obtained separately from the real part of the dielectric susceptibility) and the relaxation time $\tau_d$ as free parameters to describe the dielectric spectrum.

Fig. 1 shows the measurement \cite{tina} of the dielectric spectrum in the vacuum pump oil PPE at 255 K, where the shear spectrum \cite{tina} measured at the same temperature is perfectly fitted with $\beta=0.484$ and $\tau_c=0.209$ s. Inserting these two values, one can make two fits with eq. (\ref{ld}), the one with the $\delta$-function (the continuous line in Fig. 1), the other without (the dashed line). The necessity of a small Debye peak at the maximum is clearly demonstrated in the magnified peak region, in the insert of Fig. 1.

One can do the determination a bit differently, in order to check whether the theoretical prediction provides the correct amplitude of the $\delta$-peak. To do this, one leaves the amplitude $a$ of the $\delta$-peak as a third free parameter. For the five measurements between 250 K and 260 K in PPE \cite{tina}, the eight measurements between 214 K and 228 K in DC704, and the seven measurements between 254 K and 266 K in triphenylethylen \cite{nibo}, one obtains the average values in Table I, which agree within their error bars with the prediction of eq. (\ref{a}). The results show that even for a weak dielectric like triphenylethylen, with a rather short $\tau_d$, one still can get an indication of a small $\delta$-function on top of its dielectric relaxation peak, proving the validity of the shear misfit model.

\begin{table}[htbp]
	\centering
		\begin{tabular}{|c|c|c|c|c|}
\hline
          substance    & $\Delta\epsilon$   & $\tau_d/\tau_c$ & $a$             &  $\exp(-2\tau_c/\tau_d)$  \\
\hline
                       &                    &                 &                 &                           \\
\hline
   PPE                 & 1.82               &    0.89$\pm$0.02& 0.104$\pm$0.006 &      0.106                \\
   DC704               & 0.31               &    0.74$\pm$0.02& 0.048$\pm$0.019 &      0.066                \\
   TPE                 & 0.045              &    0.46$\pm$0.02& 0.015$\pm$0.011 &      0.013                \\
\hline
		\end{tabular}
	\caption{Average values of the dielectric strength $\Delta\epsilon$, the relaxation time ratio $\tau_d/\tau_c$, and the $\delta$-function amplitude $a$ for five measurements at different temperatures in PPE \cite{tina}, eight measurements at different temperatures in DC704 \cite{tina}, and seven measurements at different temperatures in TPE (triphenylethylene) \cite{nibo}. The measured $a$-values agree within their error bars with the prediction from eq. (\ref{a}) in the last column.}
	\label{tab:rse}
\end{table}

In rhe monoalcohols \cite{mono}, one has the special case of a very long lifetime $\tau_d$ of the local dipoles, one or two decades longer than the end of the shear relaxation at the Eshelby lifetime $\tau_c$, so according to eq. (\ref{a}) the Debye peak at $\tau_d$ dominates the whole dielectric spectrum.   

The validity of eq. (\ref{ld}) (with $a=0$, which gives a slightly better fit, though the data are not accurate enough to decide) for the description of the dynamic adiabatic compressibility is demonstrated in Fig. 2 for the two vacuum pump oils DC704 and PPE, where there is no secondary relaxation peak, and again the adiabatic compressibility data \cite{tina} are measured in the same cryostat and for the same sample as shear relaxation data. Table II compiles the fit results for three temperatures in PPE, four temperatures in DC704, and one temperature in squalane.

In all three substances, one knows from thermal expansion measurements \cite{bo,ag2020} that the isothermal density fluctuations equilibrate with the terminal relaxation time 1.75 $\tau_c$. The $\tau_d$-values for the three substances are a factor of two to four shorter, showing that the adiabatic compressibility equilibrates much earlier than the isothermal one. Since the reversible spectrum ends at $\tau_c$, and the irreversible spectrum begins there, this implies that the adiabatic compressibility is practically only due to the reaction of the soft vibrational modes \cite{ramos,schober,bouch} in the reversible Eshelby relaxations.

The adiabatic compressibility measurement determines the short time modulus $B$ and the adiabatic compressibility jump $\Delta\kappa_{ad}$. Having these, and the short time shear modulus $G$ from the shear relaxation measurement, one can calculate $\Delta\kappa_{iso}$ from eq. (\ref{dkap}). In all eight cases in Table II, $\Delta\kappa_{iso}$ is never more than thirty percent larger than $\Delta\kappa_{ad}$, showing that the larger part of the density equilibration occurs in the reversible Eshelby relaxations.

\begin{figure}[b]
\hspace{-0cm} \vspace{0cm} \epsfig{file=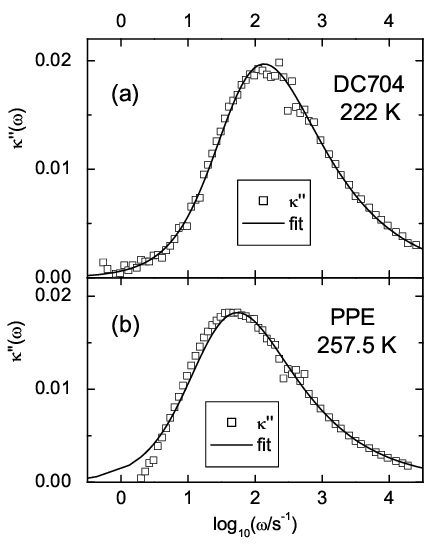,width=6 cm,angle=0} \vspace{0cm} \caption{Fit of dynamic adiabatic compressibility data (a) at 218 K in DC704 \cite{tina} (b) at 257.5 K in PPE \cite{tina} in terms of the spectrum of eq. (\ref{ld}) (continuous line).}
\end{figure}

\begin{table}[htbp]
	\centering
		\begin{tabular}{|c|c|c|c|c|c|c|c|c|}
\hline
          subst        & $T$ & $G$ & $\tau_c$ & $\tau_d/\tau_c$ & $B$  & $\Delta\kappa_{ad}$ &$\Delta\kappa_{iso}$& $B_l$  \\
\hline
                       &  K  & GPa &    s     &                 & GPa  &     GPa$^{-1}$      &GPa$^{-1}$          &  GPa   \\
\hline
   PPE                 & 255 & 0.94& 0.209    &    0.62         & 5.9  &      0.057          & 0.069              &  4.2   \\
                       &257.5& 0.91& 0.045    &    0.64         & 5.7  &      0.061          & 0.074              &  4.0   \\
                       &260  & 0.87& 0.010    &    0.67         & 5.5  &      0.061          & 0.074              &  3.9   \\
\hline                          
   DC704               &218  & 1.02& 0.292    &    0.72         & 4.9  &      0.065          & 0.083              &  3.5   \\
                       &220  & 0.98& 0.071    &    0.65         & 4.8  &      0.065          & 0.083              &  3.4   \\
                       &222  & 0.96& 0.018    &    0.65         & 4.7  &      0.066          & 0.084              &  3.4   \\
                       &224  & 0.93& 0.005    &    0.66         & 4.6  &      0.066          & 0.084              &  3.3   \\
\hline                       
   SQ                  &171.65&1.1 & 4.66     &    0.29         & 5.2  &      0.076          & 0.098              &  3.3   \\
\hline
		\end{tabular}
	\caption{Fit results from shear relaxation and adiabatic compressibility in PPE, DC704, and SQ (squalane). $\Delta\kappa_{iso}$ is determined from $\Delta\kappa_{ad}$ via eq. (\ref{dkap}), which together with the $B$-value in column 5 allows to calculate the liquid bulk modulus $B_l$ in the last column. For the role of the secondary relaxation peak in the squalane measurement see text.}
	\label{tab:rs}
\end{table}

In the third example squalane, one has to deal with a strong secondary relaxation peak. The new shear relaxation data in squalane \cite{squa2017} include the measurement of the secondary relaxation peak at 148 K, about twenty degrees below the glass temperature, allowing for a fit of its parameters without any disturbing influence of the flow process. Previous investigations of molecular glass formers in the glass phase with the much more powerful dielectric spectroscopy \cite{gainaru} have demonstrated that secondary relaxation peaks are well described in terms of a gaussian in thermal activation barriers, with a maximum barrier $V_G$ and a full width at half maximum which is about $V_G/2$. Fig. 3 shows that, for a microscopic $\tau_0=10^{-13}$ s, the data at 148 K in squalane \cite{squa2017} are well described by $V_G=0.264$ eV and a full width at half maximum of $0.53V_G$, close to the values $V_G=0.27$ eV and the full width at half maximum of $0.45V_G$ of the earlier fit \cite{bu2018b} of the old data \cite{nibo}. The value found for $V_G$ can be understood quantitatively in terms of a Helfand crankshaft motion \cite{helfand} in the polymer chain, as explained in detail in reference \cite{bu2018b}. 

\begin{figure}[b]
\hspace{-0cm} \vspace{0cm} \epsfig{file=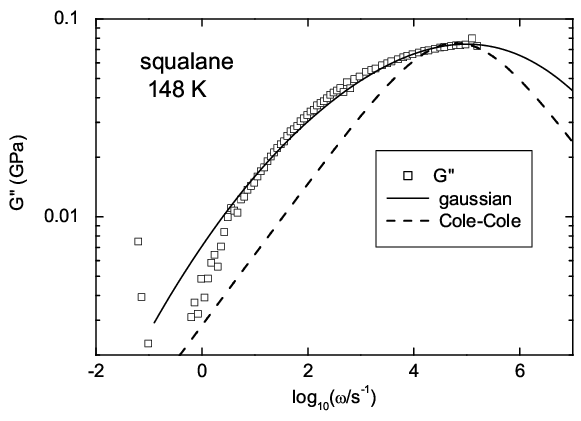,width=6 cm,angle=0} \vspace{0cm} \caption{Secondary relaxation peak in squalane in the glass phase \cite{squa2017}, together with fits in terms of a Cole-Cole function \cite{squa2017} and in terms of the gaussian barrier distribution of the present paper.}
\end{figure}

The gaussian description is obviously better than the Cole-Cole function used in the electric-circuit equivalent model to fit the shear relaxation data \cite{squa2017}, and has the additional advantage that the two parameters $V_G$ and full width at half maximum can be considered to be temperature-independent. With this assumption, and a temperature-independent Kohlrausch $\beta$ of 1/2, both sets of shear relaxation data \cite{nibo,squa2017} were fitted again in terms of the shear misfit model. Table III lists the four free parameters $G$, $\tau_c$, $a_G$, and $f_K$ of the new fit for the old data \cite{nibo}. For the new data \cite{squa2017}, the parameters were the same within experimental error, with the only difference that the $G$-values of the old data were a factor of 1.3 larger than the ones for the new data. 

\begin{figure}[t]
\hspace{-0cm} \vspace{0cm} \epsfig{file=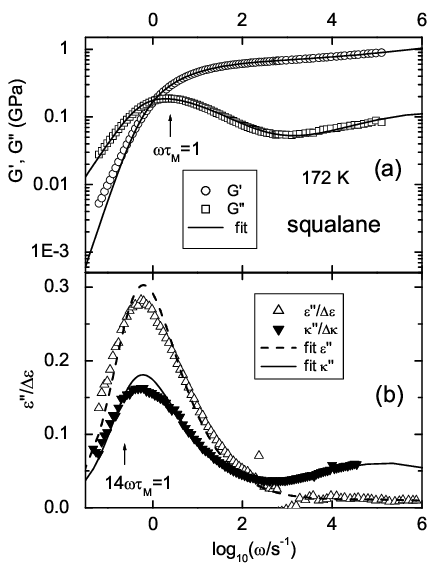,width=6 cm,angle=0} \vspace{0cm} \caption{(a) Measurement \cite{nibo} of $G(\omega)$ in squalane; fit in terms of eq. (\ref{jom}) with the parameters in Table III. (b) Fit of dielectric data for the same sample in the same cryostat \cite{nibo} with eq. (\ref{ld}), with a factor of 0.4 weaker secondary relaxation peak; fit of new adiabatic compressibility data \cite{hex} at the slightly lower temperature 171.65 K in terms of eq. (\ref{ld}) with $a=0$ and a factor 2.6 stronger secondary relaxation peak than the shear data (see text).}
\end{figure}

Fig. 4 (a) shows the older shear data of squalane \cite{nibo} at 172 K. The line is a fit in terms of eq. (\ref{jom}) with the parameters $G$, $\tau_c$, $a_G$, and $f_K$ in Table III. Taking $f_K$ as free parameter, one finds eq. (\ref{fks}) with $f_s=0.9$ reasonably well confirmed (see the $f_s$-values in column 7 of Table III). Fig. 4 (b) displays dielectric relaxation data taken on the same sample in the same cryostat \cite{nibo}, which show the secondary relaxation at the same position and with the same width as the shear data, but with an amplitude which is a factor 0.18 smaller, qualitatively consistent with the Helfand crankshaft motion \cite{helfand}, which only turns one C-C-bond around, but causes a very sizable local shear distortion. The dielectric data can be fitted with the shear parameters in Table III, with $\tau_d=1.02\tau_c$, but it turns out that this is no longer an exponential cutoff of the spectrum. The dashed line in Fig. 4 (b) requires a Kohlrausch cutoff, with a Kohlrausch $\beta_d=0.55$ at 172 K. Fitting at different temperatures, one finds that the broadening increases as one approaches the $\alpha-\beta$-merging of secondary and primary peak. 

The same broadening on approaching the $\alpha-\beta$-merging is observed in very recent adiabatic compressibility data \cite{hex}. Fig. 4 (b) shows these data at the slightly lower temperature 171.65 K, in which the Kohlrausch broadening exponent is $\beta_d=0.6$ and the secondary relaxation peak is by a factor 1.77 stronger than in the shear data.

\begin{figure}[b]
\hspace{-0cm} \vspace{0cm} \epsfig{file=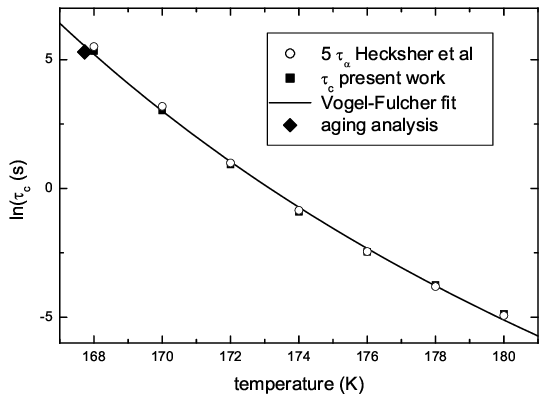,width=6 cm,angle=0} \vspace{0cm} \caption{Vogel-Fulcher fit of the values $5\tau_\alpha$ from the shear relaxation data \cite{squa2017} in the electric-circuit equivalent model (open circles), and of the $\tau_c$-values of the shear misfit model in Table III (full squares). The value $\tau_c=200$ s (full diamond) at 167.73 K is needed for the fit of the aging data in Fig. 6.}
\end{figure}

\begin{table}[htbp]
	\centering
		\begin{tabular}{|c|c|c|c|c|c|c|c|}
\hline
            $T$    & $G$   &$\tau_c$& $a_G$ &  $f_K$  & $I$  & $f_s$ & $GJ_0$     \\
\hline
            K      & GPa   &   s    &       &         &      &       &            \\
\hline
           168     & 1.38  & 208.6  & 0.067 & 1.60    & 0.67 & 0.90  &  4.17      \\
           170     & 1.42  & 18.9   & 0.088 & 1.80    & 0.89 & 0.90  &  4.71      \\
           172     & 1.45  & 2.25   & 0.111 & 1.96    & 1.11 & 0.86  &  5.17      \\
           174     & 1.46  & 0.35   & 0.130 & 2.11    & 1.28 & 0.87  &  5.58      \\
           176     & 1.42  & 0.075  & 0.139 & 2.23    & 1.35 & 0.91  &  5.83      \\
           178     & 1.35  & 0.0205 & 0.139 & 2.27    & 1.32 & 0.96  &  5.87      \\
           180     & 1.21  & 0.0072 & 0.121 & 2.32    & 1.13 & 1.13  &  5.75      \\
\hline
		\end{tabular}
	\caption{Temperature-dependent parameters of the shear misfit model for shear relaxation data \cite{nibo} in squalane, significance see text. The Kohlrausch $\beta=1/2$, the maximum secondary relaxation barrier $V_G=0.264$ eV and the full width at half maximum of the secondary relaxation gaussian 0.53 $V_G$ are fixed.}
	\label{tab:rse1}
\end{table}

Looking at the low frequency end of Fig. 4 (a), one finds a deviation of the measured $G'(\omega)$-values from the calculated curve toward higher values. This is not a failure of the shear misfit model, but a polymer effect \cite{mcleish}: Longer polymers develop a rubbery plateau, together with a much higher viscosity, due to chain entanglements, in this region. The short-chain polymer squalane shows only a small precursor effect of this rubbery plateau, but it is naturally an effect which is not taken into account here. In order to minimize the influence of the rubbery plateau precursor effect on the fitted parameters, the fit was only extended down to about one third of the peak maximum in $G''(\omega)$.

Fig. 5 shows that the fitted values of $\tau_c$ in Table III follow the Vogel-Fulcher law
\begin{equation}
	\ln{\tau_c}=\frac{1318}{T-134.4}-34.
\end{equation}

The $\tau_\alpha$-values of the electrical-circuit equivalent model evaluated from the new squalane shear relaxation data \cite{squa2017} correspond to the $\tau_c$ of the shear model. From fit of the same data in terms of the the shear misfit model, they have to be multiplied with the factor 5 to obtain $\tau_c$.

\subsection{Terminal stage measurement in squalane}

The very recent beautiful high precision aging measurement in squalane \cite{ag2020} demonstrates beyond any possible doubt the existence of a terminal stage of the glassy relaxation. The measurement determines the change of the capacitance of a planar squalane sample between two metal plates after a small temperature change as a function of time.

Squalane is a happy choice for this kind of measurement, because its low frequency dielectric constant is only $\Delta\epsilon=0.014$ higher than its high frequency dielectric constant 2.15. Therefore the temperature dependence of its dielectric constant is negligible, and the change of the capacitance reflects exclusively the density change.

Fig. 6 shows the aging data, obtained at 167.73 K, where the terminal stage of the aging lies between ten and several thousand seconds. The down triangles show the density decrease on cooling down from a 0.06 K higher temperature, the up triangles the density increase (but with the opposite sign to demonstrate the near identity) on heating from a 0.06 K lower temperature. The equilibrium density decay function at 167.73 K must lie between these two data point groups.

If one adapts $\tau_c=200$ s, the dashed curve calculated from the irreversible spectrum of eq. (\ref{pt}) does indeed fall between the data down to 100 s, but then levels off to a constant value, leaving the smaller earlier part of the total decay unexplained. The continuous line in Fig. 6, which provides a good fit of the whole measured curve, is obtained by adding the decay from the adiabatic spectrum in Fig. 4 (b) with an appropriate factor. This fit ascribes a fraction 0.3 of the total decay to the adiabatic spectrum, the fraction 0.7 to the irreversible processes.

\begin{figure}[t]
\hspace{-0cm} \vspace{0cm} \epsfig{file=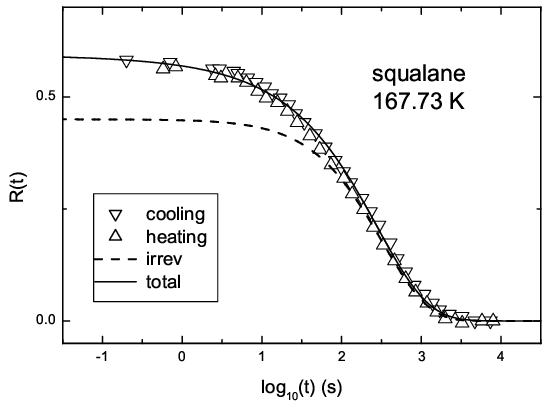,width=6 cm,angle=0} \vspace{0cm} \caption{(a) Measurement \cite{ag2020} of the density decrease of squalane at 167.73 K on cooling from a 0.06 K higher temperature (down triangles), and of the density increase on heating from a 0.06 lower temperature (up triangles), plotted with the opposite sign to demonstrate the near equality of both curves. The theoretical predictions are calculated for $\tau_c=200$ s, the dashed curve only for the irreversible spectrum of eq. (\ref{pt}), the continuous curve adds the adiabatic spectrum fitted to the data in Fig. 4 (b) with an appropriate factor (see text).}
\end{figure}

\section{Discussion and conclusions}

\subsection{Comparison of the two shear relaxation models}

The two models applied to the shear relaxation data of squalane \cite{nibo,squa2017}, the electric-circuit equivalent model \cite{squa2017} (identifying the voltage with the shear stress and the current with the shear compliance) and the shear misfit model described here, have much in common. Both are models for the time-dependent shear compliance, both assume a Kohlrausch $\beta$ of 1/2 with a cutoff at low frequency, and both describe the secondary relaxation peak as an independent feature.

An even deeper common basis is that both models are electric-circuit equivalent models, because a reversible $l(v)dv$ corresponds to a resistor in series with a capacitance, and the integral over $v$ corresponds to a combination of all these elements in parallel. The combination of all irreversible Eshelby relaxations of the shear misfit model gives rise to the viscous flow, the single resistor of the electric-circuit equivalent model. The consideration shows that many different electric-circuit models are possible; the one chosen in reference \cite{squa2017} is not unique. 

The difference lies not only in the different function for the secondary relaxation peak, but also in a different cutoff for the Kohlrausch barrier density at high barriers, a sharp cutoff in the electric-circuit equivalent model and a cutoff proportional to $\exp(-(1/\omega\tau_c)^0.8)$ in the shear misfit model. The sharp cutoff leads to the cutoff relaxation time $\tau_\alpha$ which is a factor of five lower than the $\tau_c$ of the shear misfit model. 

Naturally, shear relaxation data are not the best information source for the cutoff, because the decrease of the reversible contribution to $J''(\omega)$ with decreasing $\omega$ is overcompensated by the increase of the viscous contribution.

Obviously, the shear relaxation data allow one to choose between an infinite number of possible electric-circuit equivalent models. To decide between them, one needs additional information, like the one provided by the distinction of reversible and irreversible processes in the shear misfit model.

\subsection{Irreversible jumps and terminal stage}

The value of the concept of irreversible relaxations, the basis of the shear misfit model, lies in their different contributions to the shear compliance and to the time-dependent heat capacity or density. Under a shear stress, they contribute again and again to the viscous flow, without end. But after a small temperature change, they stop their contribution to the heat content or density change when the sample equilibrates.

Therefore their spectrum, of which only the integral enters into the viscosity, becomes visible as part of the terminal stage in the dynamic heat capacity, and in the time-dependent density change of Fig. 6. The fact that this spectrum describes the terminal stage with high accuracy, with a $\tau_c$ which agrees within the error bars with the one extrapolated from shear relaxation data, provides a convincing proof for its theoretical explanation in terms of irreversible Eshelby relaxations in the five-dimensional shear space \cite{bu2018a}.

Both dynamic heat capacity and dynamic thermal expansion contain the irreversible spectrum of eq. (\ref{pt}), but they differ in the contribution from the reversible Eshelby relaxations. The terminal expansion data \cite{ag2020} of squalane in Fig. 6 are well fitted in terms of a coupling factor for not only irreversible relaxations, but also the reversible relaxations which dominate the adiabatic compressibility. The fact that the reversible relaxations have a smaller contribution to the dynamic heat capacity data than to the dynamic thermal expansion becomes visible in their slightly longer relaxation times (factor about 1.15) in DC704 and PPE \cite{bo}.

One should be aware that the terminal stage of the shear misfit model is often, but not always the terminal stage of the highly viscous flow. It is rather the terminal stage of the structural energy and density equilibration. In this role, it also appears in soft matter, like polymers or rubbers, where the relaxation curves after a temperature step look quite similar to Fig. 6, and even the shear relaxation looks similar to Fig. 4 (a) - only it does not end in a viscous flow, but rather in a rubbery plateau \cite{mcleish,alexei} of the modulus, several decades lower than the glass modulus $G$. In these cases, the Maxwell time is much higher than $\tau_c$, the terminal stage is called segmental relaxation, and the glass transition occurs when the segmental relaxation time $\tau_c$ gets so long that the energy and density fluctuations remain frozen on an experimental time scale.

Something similar happens in the monoalcohols \cite{mono}, where the structural hydrogen bond connections survive the terminal stage, leading to a terminal dielectric relaxation time much longer than $\tau_c$ (see also the treatment of hydrogen-bonded glass formers in terms of the shear misfit model \cite{hb}).

\subsection{Density fluctuations}

In DC704, it is possible to show that eq. (\ref{dkap}) for the ratio of the compressibility jumps $\Delta\kappa_{ad}$ and $\Delta\kappa_{iso}$ at the glass transition is corroborated independently by other data. The difference between isothermal and adiabatic compressibility in the liquid is
\begin{equation}
	\kappa_{ad}=\frac{c_V}{c_p}\kappa_{iso},
\end{equation}
with the difference between $c_p$ and $c_V$ calculable from the thermodynamic relation
\begin{equation}
	c_p-c_V=\alpha_l^2B_lVT,
\end{equation}
where $\alpha_l$ is the thermal volume expansion coefficient, and $B_l$ is the bulk modulus of the liquid. In DC704, one knows \cite{gundermann} $c_p=1.65\ 10^6$ J/m$^3$K, $\alpha_l=4.6\ 10^{-4}$ K$^{-1}$, and $B_l=3.54$ GPa at $T_g=214$ K, so $\kappa_{ad}=0.909\kappa_{iso}$. The total isothermal $\Delta\kappa$ at the glass transition is 0.082 GPa$^{-1}$, so the measured adiabatic $\Delta\kappa=0.065$ is a factor 0.789 smaller, almost exactly the diminution factor $3B/(3B+4G)$ of eq. (\ref{dkap}) in Section II.C.

If eq. (\ref{dkap}) is indeed to be trusted, it tells us that in squalane the difference $\Delta\kappa_{iso}-\Delta\kappa_{ad}$ is only about one fifth of $\Delta\kappa_{ad}$. But in the fit of the thermal expansion data \cite{ag2020} in Fig. 6 this small minority is a factor 2.3 more prominent than the adiabatic mayority, more precisely a factor 10.5 larger than expected on the basis of an equal footing. The difference $\Delta\kappa_{iso}-\Delta\kappa_{ad}$ relaxes with the irreversible Eshelby relaxations. If it appears a factor 10.5 larger than expected, one must conclude that the reversible Eshelby relaxations are a factor 10.5 weaker than the irreversible ones in experiments involving a temperature step.

To understand this, remember the difference between structural excess entropy and vibrational excess entropy of the undercooled liquid over the glass. The larger structural excess entropy equilibrates in the irreversible Eshelby transitions. But since the adaptation of the density occurs already in the reversible Eshelby relaxations, they should be able to adapt their vibrational excess entropy, due to the soft vibrational modes which are responsible for the low temperature glass anomalies \cite{ramos,schober,bouch}.

To get a quantitative estimate, remember that a neutron scattering measurement of the vibrational excess entropy in selenium \cite{se} determined a fraction of 0.28 of the total excess entropy. The reversible Eshelby relaxations in squalane with its strong secondary relaxation peak are about one third of the total, so our estimate for the fraction of the total excess entropy equilibrated by the reversible Eshelby relaxations ends up close to the 1/10.5 inferred from the analysis of the terminal stage data \cite{ag2020} in Fig. 6.

The consideration is able to solve another old glass transition puzzle, the deviation of the Prigogine-Defay ratio
\begin{equation}\label{prigo}
	\Pi=\frac{\Delta c_p\Delta\kappa_{iso}}{(\Delta\alpha)^2T}
\end{equation}
from one \cite{jackle}, where $\Delta c_p$ and $\Delta\alpha$ are the heat capacity and thermal expansion jumps at the glass transition, respectively. The most dramatic case is vitreous silica \cite{dingwell}, where the Prigogine-Defay ratio is close to infinity, because the change $\Delta\alpha$ of the thermal expansion between glass and liquid is close to zero. In this case, the soft modes have a strong negative Gr\"uneisen parameter, and are responsible for the the negative expansion coefficient at low temperatures \cite{barron}. Above 150 K, higher frequency modes with positive Gr\"uneisen parameters take over and lead to a small positive thermal expansion.

At the glass temperature of silica, a temperature increase leads to a pronounced increase of the soft mode density, as evidenced by measurements of the low temperature anomalies in samples annealed at different temperatures \cite{lohneysen}). This increase in the number of soft modes leads to a contraction of the sample, a negative contribution to $\Delta\alpha$, but contributes positively to both $\Delta c_p$ and to $\Delta\kappa_{iso}$. In the terminal irreversible processes, there is obviously also some energy input into the higher degrees of freedom, leading to an expansion which in silica happens to compensate the soft mode contraction. In total, the resulting thermal expansion of the liquid is practically equal to the one of the glass, making the Prigogine-Defay-ratio extremely high \cite{dingwell}.

One concludes that the physical reason for the deviation of the Prigogine-Defay-ratio from one \cite{jackle} is the deviation of the Gr\"uneisen parameter of the soft modes behind the low temperature glass anomalies \cite{ramos,schober,bouch} from the average one.

\subsection{Conclusions}

The shear misfit model, describing the terminal stage of the highly viscous flow in terms of irreversible Eshelby or shear transformation zone relaxations, and the initial stage in terms of a Kohlrausch tail of reversible Eshelby shear relaxations, has been reformulated in a critical survey, also for the case of an additional secondary relaxation peak. The existence of a small Debye peak in dielectric data has been derived, together with a relation between adiabatic and isothermal compressibility changes at the glass transition. In the model, the crossover from irreversible to reversible Eshelby processes occurs at an Eshelby region lifetime which is a factor of eight longer than the Maxwell time.

The model is applied to literature measurements of shear, dielectric, bulk, and aging relaxation in two vacuum pump oils and squalane, a short chain polymer with a strong secondary relaxation peak. The aging measurement shows the terminal stage of the highly viscous flow with unprecedented precision. The shear misfit model is able to reproduce this precise measurement with the lifetime extrapolated from the fit of the shear relaxation data. The findings demonstrate not only the validity of the irreversible Eshelby relaxation approach, but also the one of its Kohlrausch extension to reversible Eshelby processes, both essential elements of a future exact theory of highly viscous liquids.


\begin{thebibliography}{99}
\bibitem{jackle} J. J\"ackle, J. Chem. Phys. {\bf 79}, 4463 (1983)
\bibitem{cavagna} A. Cavagna, Phys. Rep. {\bf 476}, 51 (2009)
\bibitem{bb} L. Berthier and G. Biroli, Rev. Mod. Phys. {\bf 83}, 587 (2011)
\bibitem{royall} C. P. Royall and S. R. Williams, Phys. Rep. {\bf 560}, 1 (2015)
\bibitem{berthier} L. Berthier, J. Chem. Phys. {\bf 150}, 160902 (2019)
\bibitem{shov} J. C. Dyre, N. B. Olsen, and T. Christensen, Phys. Rev. B {\bf 53}, 2171 (1996)
\bibitem{falk} M. L. Falk and J. S. Langer, Phys. Rev. E {\bf 57}, 7192 (1998)
\bibitem{johnson} W. L. Johnson and K. Samwer, Phys. Rev. Lett. {\bf 95}, 195501 (2005)
\bibitem{ngai} K. L. Ngai, {\it Relaxation and Diffusion in Complex Systems} (Springer, New York 2011)
\bibitem{dyre} J. C. Dyre, Rev. Mod. Phys. {\bf 78}, 953 (2006)
\bibitem{random} T. B. Schroeder and J. C. Dyre, Phys. Chem. Chem. Phys. {\bf 4}, 3173 (2002)
\bibitem{bu2018a} U. Buchenau, J. Chem. Phys. {\bf 148}, 064502 (2018)
\bibitem{bu2018b} U. Buchenau, J. Chem. Phys. {\bf 149}, 044508 (2018)
\bibitem{ramos} M. A. Ramos and U. Buchenau, Phys. Rev. B {\bf 55}, 5749 (1997)
\bibitem{schober} D. A. Parshin, H. R. Schober, and V. L. Gurevich, Phys. Rev. B {\bf 76}, 064206 (2007)
\bibitem{bouch} E. Lerner, and E. Bouchbinder, J. Chem.Phys. {\bf 155}, 200901 (2021)
\bibitem{eshelby} J. D. Eshelby, Proc. Roy. Soc. {\bf A241}, 376 (1957)
\bibitem{tina} T. Hecksher, N. B. Olsen, K. A. Nelson, J. C. Dyre and T. Christensen, J. Chem. Phys. {\bf 138}, 12A543 (2013)
\bibitem{nibo}  B. Jakobsen, K. Niss, and N. B. Olsen, J. Chem. Phys. {\bf 123}, 234510 (2005)
\bibitem{squa2017} T. Hecksher, N. B. Olsen, and J. C. Dyre, J. Chem. Phys. {\bf 146},154504 (2017)
\bibitem{ag2020} K. Niss, J. C. Dyre, and T. Hecksher, J. Chem. Phys. {\bf 152}, 041103 (2020)
\bibitem{hex} L. A. Roed, J. C. Dyre, K. Niss, T. Hecksher, and B. Riechers, J. Chem. Phys. {\bf 154}, 184508 (2021)
\bibitem{bnap} R. B\"ohmer, K. L. Ngai, C. A. Angell and D. J. Plazek, J. Phys. Chem. {\bf 99}, 4201 (1993)
\bibitem{albena} A. I. Nielsen, T. Christensen, B. Jakobsen, K. Niss, N. B. Olsen, R. Richert, and J. C. Dyre, J. Chem. 
Phys. {\bf 130}, 154508 (2009)
\bibitem{hb} U. Buchenau, arXiv:2105.06392
\bibitem{gainaru} C. Gainaru, R. B\"ohmer, R. Kahlau, and E. R\"ossler, Phys. Rev. B {\bf 82}, 104205 (2010)
\bibitem{corei} V. A. Luchnikov, N. N. Medvedev, Yu. J. Naberukhin, and H. R. Schober, Phys. Rev. B {\bf 62}, 3181 (2000)
\bibitem{corein} M. Shimada, H. Mizuno, M. Wyart, and A. Ikeda, Phys. Rev. E {\bf 98}, 060901 (2018) 
\bibitem{bo} B. Jakobsen, T. Hecksher, T. Christensen, N. B. Olsen, J. C. Dyre, and K. Niss, J. Chem. Phys. {\bf 136}, 081102 (2012)
\bibitem{mono} R. B\"ohmer, C. Gainaru, and R. Richert, Phys. Rep. {\bf 545}, 125 (2014)
\bibitem{helfand} E. Helfand, J. Polym. Sci., Polym. Symp. {\bf 73}, 39 (1985)
\bibitem{read} B. E. Read, Polymer {\bf 22}, 1580 (1981)
\bibitem{mcleish} T. C. B. McLeish, Adv. Phys. {\bf 51}, 1379 (2002)
\bibitem{alexei} A. L. Agapov, V. N. Novikov, T. Hong, F. Fan, and A. P. Sokolov, Macromolecules {\bf 51}, 4874 (2018)
\bibitem{gundermann} D. Gundermann, U. R. Pedersen, T. Hecksher, N. P. Bailey, B. Jakobsen, T. Christensen, N. B. Olsen, Th. B. Schroeder, D. Fragiadakis, R. Casalini, C. M. Roland, J. C. Dyre, and K. Niss, Nature Physics {\bf 7}, 816 (2011)
\bibitem{se} W. A. Phillips, U. Buchenau, N. N\"ucker, A. J. Dianoux, and W. Petry, Phys. Rev. Lett. {\bf 63}, 2381 (1989)
\bibitem{dingwell} D.B. Dingwell, R. Knochc, and S.L. Webb, Phys. Chem. Minerals {\bf 19}, 445 (1993)
\bibitem{barron} T. H. K. Barron, J. G. Collins, and G. K. White, Adv. Physics {\bf 29}, 609 (1980)
\bibitem{lohneysen} H. v. L\"ohneysen, H. R\"using, and W. Sander, Z. Phys. B - Condensed Matter {\bf 60}, 323 (1985)
\end{thebibliography}
\end{document}